# Variational Autoencoders with a Structural Similarity Loss in Time of Flight MRAs

Kimberley M. Timmins*[a], Irene C. van der Schaaf[b], Ynte M. Ruigrok[c], Birgitta K. Velthuis[b], Hugo J. Kuijf[a].

[a]Image Sciences Institute, University Medical Center Utrecht, Utrecht, The Netherlands;
[b]Department of Radiology, University Medical Center Utrecht, Utrecht, The Netherlands;
[c]Department of Neurology and Neurosurgery, UMC Utrecht Brain Center, University Medical Center Utrecht, Utrecht, The Netherlands

*kmtimmins@umcutrecht.nl, +31 (0) 88 75 59863

## ABSTRACT

Time-of-Flight Magnetic Resonance Angiographs (TOF-MRAs) enable visualization and analysis of cerebral arteries. This analysis may indicate normal variation of the configuration of the cerebrovascular system or vessel abnormalities, such as aneurysms. A model would be useful to represent normal cerebrovascular structure and variabilities in a healthy population and to differentiate from abnormalities. Current anomaly detection using autoencoding convolutional neural networks usually use a voxelwise mean-error for optimization. We propose optimizing a variational-autoencoder (VAE) with structural similarity loss (SSIM) for TOF-MRA reconstruction.

A patch-trained 2D fully-convolutional VAE was optimized for TOF-MRA reconstruction by comparing vessel segmentations of original and reconstructed MRAs. The method was trained and tested on two datasets: the IXI dataset, and a subset from the ADAM challenge. Both trained networks were tested on a dataset including subjects with aneurysms. We compared VAE optimization with L2-loss and SSIM-loss. Performance was evaluated between original and reconstructed MRAs using mean square error, mean-SSIM, peak-signal-to-noise-ratio and dice similarity index (DSI) of segmented vessels.

The L2-optimized VAE outperforms SSIM, with improved reconstruction metrics and DSIs for both datasets. Optimization using SSIM performed best for visual image quality, but with discrepancy in quantitative reconstruction and vascular segmentation. The IXI dataset had overall better performance, potentially due to the larger, more diverse training data. Reconstruction metrics, including SSIM, were lower for MRAs including aneurysms.

A SSIM-optimized VAE improved the visual perceptive image quality of TOF-MRA reconstructions. A L2-optimized VAE performed best for TOF-MRA reconstruction, where the vascular segmentation is important. SSIM is a potential metric for anomaly detection of MRAs.

## 1. INTRODUCTION

### 1.1 Background

Time-of-Flight Magnetic Resonance Angiographs (TOF-MRA) allow the visualization and analysis of the configuration of the cerebral arteries. This analysis can indicate normal variation in the cerebrovascular system or vessel abnormalities such as aneurysms or stenosis. In addition, variation in the geometry and configuration of the Circle of Willis, could indicate patients at risk for development of an aneurysm and cerebral vascular disease[1–3]. To diagnose these differences, a model would be helpful which can represent the normal variation of the cerebral vessels in a healthy population.

Previous work has demonstrated that autoencoders and variational autoencoders[4] (VAEs) can be trained on images of healthy subjects to reconstruct healthy images. When these models are presented with a different image, containing an anomaly, the model will ignore the anomaly and reconstruct the image as if it would be, as a healthy image. The anomaly or variation may be determined by comparing the reconstructed 'healthy' image with the original image. This enables autoencoders and VAEs to be used for unsupervised anomaly detection, including pathology variation in brain MRIs[5,6]. Most previously implemented VAEs for anomaly detection use per-pixel loss functions such as L1-loss or L2-loss. Such loss functions, make the assumption that intensity values of neighboring pixels are independent. These approaches are less suitable where the anomaly may result in a change in structure, rather than pixel intensity. This is true in the case for example of aneurysms or vessel irregularities, where the irregularity often has the same intensity as the surrounding vasculature in the MRA and can only be defined as an anomaly by its structure. A contextual loss would allow for more structural features in the image to be taken into consideration. The Structural Similarity Index (SSIM)[7] is an image quality measure which can identify difference in structure of images by comparing patterns in intensities which are normalized for luminance and contrast.

### 1.2 Aim

We investigate the use of the Structural Similarity Index as a loss function for optimization of a VAE for reconstructing normal TOF-MRAs in healthy patients, compared to a voxelwise L2 loss function. The experiments will be trained on a large publicly available dataset, and a smaller in-house dataset. Prediction of the trained methods will also be performed on a third dataset including subjects with aneurysms.

## 2. MATERIALS AND METHODS

### 2.1 Datasets

The IXI dataset (a) consists of 570 TOF-MRAs of healthy patients collected from three different hospitals in London, United Kingdom, between 2005 and 2006: Hammersmith Hospital (Philips Healthcare, 3T), Guy's Hospital (Philips Healthcare, 3T) and Institute of Psychiatry (GE, 1.5T). This set was randomly split into sets for training (365 MRAs), validation (91 MRAs) and test (114 MRAs).[8] The IXI database contains subjects who have been screened by radiologists and diagnosed to be healthy, and therefore the scans do not contain any diagnosed aneurysms.

The in-house dataset (b) was a subset of the healthy patient data used for the Aneurysm Detection and segMentation (ADAM) challenge for MICCAI 2020 (https://adam.isi.uu.nl/), consisting of MRAs without aneurysms collected from the University Medical Center Utrecht, the Netherlands. This consists of 46 MRAs which were randomly split into sets for training (16 MRAs), validation (3 MRAs) and test (27 MRAs). The scans were diagnosed by radiologists in the clinic as having no present un-ruptured intracranial aneurysms.

A third, additional test dataset (c) of 30 MRAs from the ADAM challenge,[9] of subjects diagnosed with un-ruptured intracranial aneurysms, was included for testing of both the trained methods. This dataset contained images using the same protocols as the in-house dataset (b). It was only used for testing and not used for training.

### 2.2 Pre-processing

All MRAs were corrected for bias field in-homogeneities with the N4 bias field correction algorithm[10]. The intensity of the images was normalized between 0 and 1, based on 1 being 95% of the maximum intensity of the original MRA. Otsu thresholding[11] was used to form a crude brain mask. Patches of 32 x 32 voxels were randomly selected from the masked, normalized MRAs in the training set. A patch size of 32 x 32 voxels was chosen, because this allowed the full width of a vessel to be included in a patch. For the IXI training dataset (a), 1,000 patches per MRA were extracted leading to a total of 365,000 patches for training. For the smaller healthy ADAM training dataset (b), 10,000 patches were extracted per MRA, with a total of 160,000 patches for training. More patches were taken from the ADAM dataset per image, as there were less images in the ADAM dataset than the IXI dataset. The same patches were used for all experiments.

## 2.3 Architecture

A fully convolutional VAE was developed using PyTorch to conserve spatial information in the latent space. The latent space was a multidimensional tensor of 32x4x4, the size of which was optimized by comparing visual quality of the reconstructed output. The reconstruction loss functions optimized were the voxelwise L2-loss function and a differentiable SSIM loss function. The SSIM loss was implemented with a window size of 11x11 and given a weighting of 1,000. The Kullback-Leiber divergence loss term was also included to standardize the distribution of latent space as is standard in VAEs. The learning rate and batch size were optimized for memory and performance with a batch size of 100 patches, and learning rate of 0.01 for L2 and 0.001 for SSIM for all experiments. A total of four networks were trained: two networks each optimizing L2 loss and SSIM loss for each of (a) the IXI and (b) the healthy ADAM training datasets. The networks were trained until convergence of the validation loss. The trained networks were then used to predict for each of the test sets: (a) the IXI healthy test set, (b) ADAM healthy test set and (c) the ADAM aneurysm test set. Since they were fully convolutional networks, the predictions were made on the full-sized original pre-processed MRAs. These were tested slice per slice before combining to the full 3D TOF-MRA. The intensity of the resulting reconstructed images was rescaled to the same intensity scale as the original image. Vessel segmentation was performed on the resulting reconstructed images using a previously trained vessel segmentation U-Net[12].

## 2.4 Evaluation

We compared the performance of two different trained networks for each dataset, one optimized with L2 loss function and the other with a SSIM loss function. Prediction was performed on the test sets from each of the same dataset, a) and b) (Table 1). Prediction of both trained networks was also performed on the test set (c) of MRAs of subjects with aneurysms (Table 2). Reconstruction performance was evaluated between the original and reconstructed images using mean square error (MSE), mean SSIM, peak signal-to-noise-ratio (PSNR). The dice similarity index (DSI) was used to determine overlap of the vessel segmentations of the reconstructed images and the original images.

# 3. RESULTS

Using L2 and SSIM loss functions for both (a) the IXI and (b) the ADAM datasets allowed for sufficient image reconstruction and vessel segmentation to be performed as shown in Table 1. Quantitative reconstruction was evaluated using MSE, mean SSIM and PSNR, with MSE close to zero and all mean SSIM > 0.7. L2 loss performed on average better than SSIM loss for all quantitative evaluative purposes, with a higher DSI, SSIM and PSNR and lower MSE for both datasets. The networks trained and tested on the IXI dataset using SSIM loss outperformed the healthy ADAM dataset with regard to quantitative reconstruction of the images when assessing with MSE, mean SSIM and PSNR. However, the MSE, mean SSIM and PSNR of the L2 trained model for ADAM were higher than the SSIM trained IXI model. All DSI scores larger than 0.5 for all reconstructed images as seen in Table 1. However, all the DSI scores were lower for the IXI dataset.

Table 1. Reconstruction and segmentation metrics for the test set for the datasets from a) the IXI dataset and b) the ADAM challenge, trained on the respective training dataset. Values are provided as mean (standard deviation). DSI: Dice similarity index of the resulting vessel segmentation, MSE: mean-square-error, SSIM: structural similarity loss, PSNR: peak-signal-to-noise-ratio, L2: L2-loss.

|  |  | **DSI** | **MSE** | **Mean SSIM** | **PSNR** |
|---|---|---|---|---|---|
| **a) IXI** | SSIM | 0.573 (0.119) | 0.003 (0.002) | 0.851 (0.039) | 26.9 (3.16) |
|  | L2 | 0.604 (0.111) | 0.001 (0.000) | 0.914 (0.024) | 33.5 (1.09) |
| **b) ADAM** | SSIM | 0.729 (0.183) | 0.012 (0.007) | 0.706 (0.089) | 20.8 (3.36) |
|  | L2 | 0.837 (0.065) | 0.001 (0.001) | 0.883 (0.031) | 29.3 (2.41) |

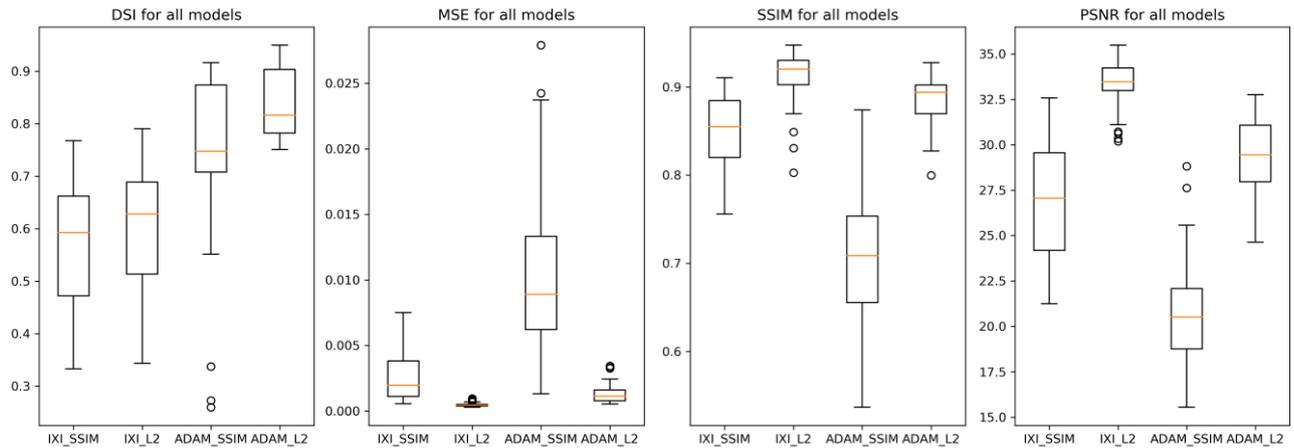

Figure 1. Box plots of reconstruction and segmentation metrics for all scans in the test set for the datasets from a) the IXI dataset and b) the ADAM challenge, trained on the respective training dataset. The center bars correspond to the median value.

Optimization of the network using SSIM loss resulted in reconstructed images with an improved visual perceptual image quality, with more structural details. These structural details were smoother on reconstruction based on the L2 loss VAE, as seen in Figure 2.

(a)          (b)          (c)

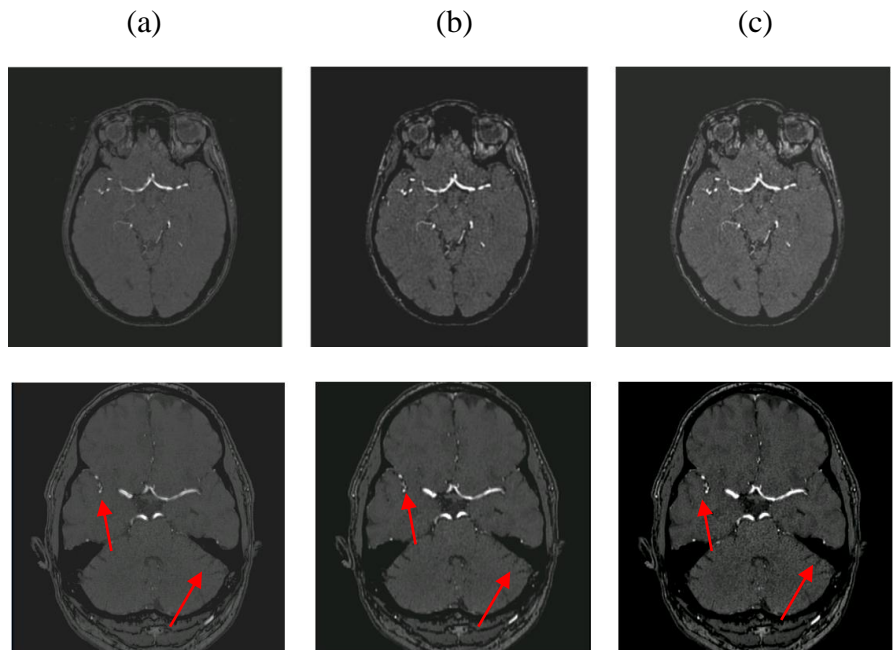

Figure 2. Reconstructions of TOF-MRAs using fully convolutional VAE trained with different loss functions a) Original TOF-MRA b) Reconstructed MRA: VAE trained with L2 loss c) Reconstructed MRA: VAE trained with SSIM loss.
Top Row: IXI healthy dataset, Bottom Row: ADAM healthy dataset
Red arrows indicate areas of more clear structure in both vessel and brain tissue in SSIM reconstructed image compared to original and L2 reconstructed image.

For (c) the third test set containing aneurysms, reconstruction metrics were poorer for all trained networks as seen in Table 2 and Figure 3. The IXI trained method had a worse DSI score and lower MSE, SSIM and PSNR than the ADAM trained method for evaluation on (c). The IXI trained VAEs had a larger DSI for (c) the aneurysm dataset compared to (a) the IXI test set.

Table 2. Reconstruction and segmentation metrics for the prediction of the third test set containing aneurysms, for each of the networks trained on a) the IXI dataset and b) the ADAM challenge dataset. Values are provided as mean (standard deviation).

| c) Aneurysm Test Set | | | | |
|---|---|---|---|---|
| **Network** | **DSI** | **MSE** | **Mean SSIM** | **PSNR** |
| IXI SSIM | 0.602 (0.217) | 0.018 (0.008) | 0.652 (0.081) | 18.0 (2.72) |
| IXI L2 | 0.782 (0.133) | 0.003 (0.001) | 0.845 (0.040) | 26.3 (2.31) |
| ADAM SSIM | 0.692 (0.185) | 0.012 (0.006) | 0.696 (0.074) | 20.1 (2.90) |
| ADAM L2 | 0.790 (0.014) | 0.001 (0.001) | 0.880 (0.028) | 28.7 (1.95) |

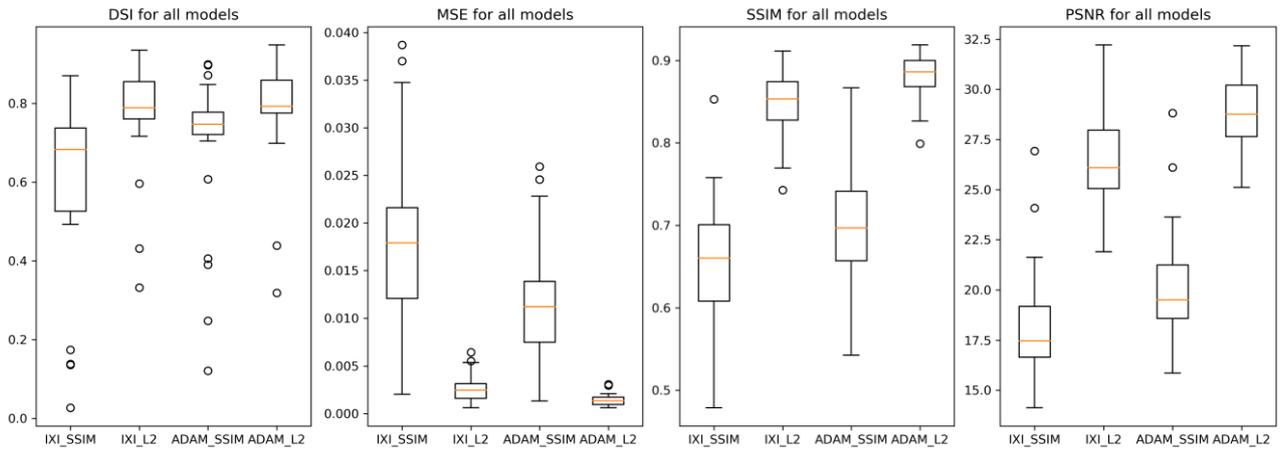

Figure 3. Box plots of reconstruction and segmentation metrics for all scans in the test set for the datasets from a) the IXI dataset and b) the ADAM challenge, trained on the respective training dataset. The center bars correspond to the median value.

## 4. DISCUSSION AND CONCLUSION

Our results show that using SSIM as a loss function in a VAE provides a better perceptual image quality than with a voxelwise L2 loss for the reconstruction of TOF-MRAs. However, L2 loss performs better for quantitative vessel segmentation and reconstruction metrics.

In the MRAs, there are multiple structures inside the brain which are emphasized when the SSIM loss is optimized (see Figure 2). This results in a lower contrast (as demonstrated by the lower PSNR relative to the L2 loss) between the vessels and surrounding brain tissue. This may result in the lower vessel segmentation performance. For vascular segmentation, these structures in the surrounding brain tissue are not of interest, but they can be for the diagnosis of other cerebral disease. For vessel segmentation from the reconstructed VAEs, selection of patches for training containing only vessel, merits further investigation as this narrows the problem. Furthermore, the vessel segmentation method was trained on original TOF-MRAs, potentially leading to a lower performance for reconstructed TOF-MRA with different image qualities compared to the original MRAs.

The higher reconstruction metrics for the IXI dataset may be caused by the larger quantity and diversity in vascular confirmation and aneurysms in the training data. The lower DSI scores of the IXI set are likely due to the fact that the vessel segmentation network was trained using data from the ADAM challenge. This results in the original vessel segmentation being suboptimal and consequently the reconstructed vessel segmentation might be even poorer. For more valid assessment of vessel segmentation performance, an alternative vessel segmentation method could be used which performs comparably for both datasets, or the current vessel segmentation network could be re-trained on the IXI dataset. A further limitation of our study, was that validation was performed on a single random split of the dataset and in future studies different validation splits or k-fold validation should be used to ensure a fair distribution of the data.

The networks trained on (a) the IXI dataset performed well on reconstruction of (c) the aneurysm dataset and were not substantially worse than the (b) healthy ADAM trained networks. The aneurysm dataset was made up of MRAs that had the same protocols and variety in field strength used in the healthy ADAM training set. This suggests that the IXI trained dataset has good performance, even on data with a different protocol from which it was trained, and could potentially work on a variety of different datasets.

Notably, the mean SSIM was lower for all trained networks on the aneurysm test set relative to the healthy test sets. This suggests that an anomaly (such as an aneurysm) may result in a reconstructed image that has a structure more similar to that of a healthy subject, less similar to the original image (lower SSIM). This can be seen in Figure 4, where the aneurysm in the reconstructed image has a lower SSIM value. This suggests that VAEs trained using either an L2 or SSIM loss may be useful for anomaly detection when evaluated against the original images using SSIM. Further investigation into this would need to be performed.

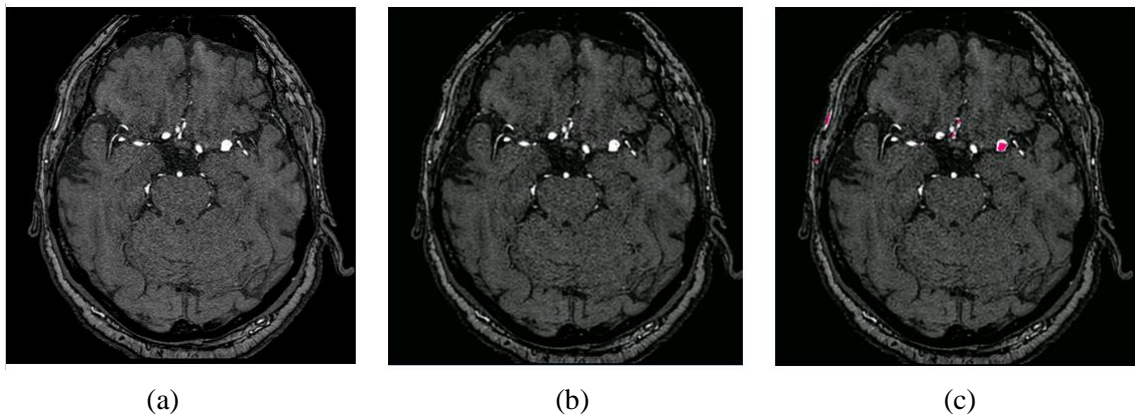

(a)         (b)         (c)

Figure 4. a) Original TOF-MRA with aneurysm b) Reconstructed using ADAM SSIM trained network c) Possible anomaly detection method using SSIM values and thresholds. Anomalies in SSIM are shown overlaid in dark pink.

In conclusion, our study demonstrates that using SSIM loss to train a VAE does improve the perceptive visual quality of the reconstructed MRA over L2 loss. However, it should be used with caution as it does not necessarily improve the quantitative voxelwise representation of specific features which may be required for future analysis. L2-loss trained VAEs may be used for accurate reconstruction of TOF-MRAs. Furthermore, we suggest that SSIM may be a potential metric for anomaly detection in TOF-MRAs.

## ACKNOWLEDGEMENTS

We acknowledge the support from the Netherlands Cardiovascular Research Initiative: An initiative with support of the Dutch Heart Foundation, CVON2015-08 ERASE and CVON2018-02 ANEURYSM@RISK.

# REFERENCES

[1] Bor, A. S. E., Velthuis, B. K., Majoie, C. B. and Rinkel, G. J. E., "Configuration of intracranial arteries and development of aneurysms: a follow-up study.," Neurology **70**(9), 700–705 (2008).

[2] Kayembe, K. N., Sasahara, M. and Hazama, F., "Cerebral aneurysms and variations in the circle of Willis.," Stroke **15**(5), 846–850 (1984).

[3] Campeau, N. G. and Huston, J., "Vascular Disorders-Magnetic Resonance Angiography: Brain Vessels," Neuroimaging Clin. N. Am. **22**(2), 207–233 (2012).

[4] Kingma, D. P. and Welling, M., "Auto-Encoding Variational Bayes," 1–14 (2013).

[5] Baur, C., Wiestler, B., Albarqouni, S. and Navab, N., "Deep autoencoding models for unsupervised anomaly segmentation in brain MR images," Lect. Notes Comput. Sci. (including Subser. Lect. Notes Artif. Intell. Lect. Notes Bioinformatics) **11383 LNCS**(April), 161–169 (2019).

[6] Zimmerer, D., Isensee, F., Petersen, J., Kohl, S. and Maier-Hein, K., "Unsupervised Anomaly Localization Using Variational Auto-Encoders," [Informatik aktuell], 289–297 (2019).

[7] Wang, Z., Bovik, A. C., Sheikh, H. R. and Simoncelli, E. P., "Image Quality Assessment: From Error Visibility to Structural Similarity," IEEE Trans. Image Process. **13**(4), 600–612 (2004).

[8] GR/S21533/02, E., "IXI Dataset – Information eXtraction from Images," <http://brain-development.org/ixi-dataset/>.

[9] UMC Utrecht, "ADAM Challenge 2020," <adam.isi.uu.nl>.

[10] Tustison, N. J., Cook, P. A. and Gee, J. C., "N4Itk," 1310–1320 (2011).

[11] Otsu, N., "A Threshold Selection Method from Gray-Level Histograms," IEEE Trans. Syst. Man. Cybern. **9**(1), 62–66 (1979).

[12] de Vos, V., Timmins, K. M., van der Schaaf, I. C., Ruigrok, Y., Velthuis, B. K. and Kuijf, H. J., "Automatic Cerebral Vessel Extraction in TOF-MRA Using Deep Learning," SPIE Med. Imaging 2021.